\documentclass[11pt]{article}
\newcommand{\be}{\begin{eqnarray}}\newcommand{\beq}{\begin{equation}}
\newcommand{\ee}{\end{eqnarray}}\newcommand{\eeq}{\end{equation}}
\newcommand{\ep}{\epsilon}\newcommand{\eps}{\varepsilon}
\newcommand{\la}{\lambda}\newcommand{\De}{\Delta}
\newcommand{\rad}{\raisebox{1.3ex}{\tiny{\ensuremath{\bullet}}}}
\usepackage{graphicx,amssymb,amsmath} 
\title{
Enthalpy effect on the kinetics of concurrent nucleation and 
chemical aging of aqueous organic aerosols: The stage of thermal relaxation 
}
\author{ 
Yuri S. Djikaev\thanks{Corresponding author. E-mail: idjikaev@buffalo.edu} 
\hspace{0.1cm} and \hspace{0.1cm}Batradz I. Djikkaity
\\ Department of Chemical and Biological  Engineering, SUNY at Buffalo,\\  
Buffalo, New York  14260 }
\date{}
\renewcommand{\baselinestretch}{1}

\begin{document}
\bibliographystyle{plain}
\renewcommand{\baselinestretch}{1}
\maketitle
\begin{abstract}
{\bf \LARGE }

The  size and composition distribution of an ensemble of aqueous organic droplets, evolving  via nucleation
and concomitant chemical aging, may be affected by the latent heat of condensation and enthalpy of
heterogeneous chemical reactions, so the temperature of the droplet may deviate from the air temperature and
thus become an independent variable of its state (additional to its size and composition variables).
Using the formalism of the classical nucleation theory, we derive  a partial differential equation for the
temporal evolution of the distribution of an ensemble of such droplets with respect to all their variables of
state   via Taylor series expansions of the corresponding multidimensional discrete equation of balance,
describing the material and  heat exchange between droplets and air.  
The resulting kinetic equation goes beyond the  framework of the Fokker-Planck approximation with respect to
the temperature variable.  A hierarchy of time scales of nonisothermal nucleation and concomitant chemical
aging of aqueous organic aerosols is established and an analytical description of their thermal  relaxation
stage is developed, allowing one to estimate the characteristic time of the establishment of the equilibrium
distribution of aerosol particles with respect to their temperatures. Theoretical results are illustrated
with  numerical calculations for the concurrent nucleation and chemical aging of model aqueous
hydrophilic-hydrophobic organic aerosols in air. 

\end{abstract}

\begin{figure}[h]
\begin{center}\vspace{0.3cm}
\vspace{-3.2cm}
\end{center}
\end{figure} 
\renewcommand{\baselinestretch}{1} 
\renewcommand{\baselinestretch}{2}
\newpage

\section{Introduction}
\normalsize
\renewcommand{\baselinestretch}{2}

Nonisothermal effects can markedly influence first-order phase transitions, particularly condensation.  
First, the heating of the nascent liquid droplets by the latent heat of condensation  causes a reduction of
the nucleation rate by increasing the ability of droplets to  emit molecules and by decreasing the
metastability of the vapor phase (due to the increase in the system  temperature).  Second, the fluctuations
of the droplet temperature exist even in the absence of matter exchange between the nuclei and the medium;
they also influence the emissivity of droplets. Third, droplets as particles of condensed matter are thermally
quasi-isolate from one another, being surrounded by the low-density vapor-gas medium. Consequently, the
temperature of a droplet decreases gradually during every event of emission of a molecule (while the molecule
passes from the nucleus through its surface layer into the vapor). Therefore, the droplet emissivity must  be
determined by some intermediate value of its temperature but neither by the initial one (before the emission
event) nor by the final one (after the emission event). Clearly, the latent heat of condensation/evaporation
can substantially affect the droplet temperature only if the density  of the carrier (passive) gas in the
system is so small  that its molecules are unable to establish  thermal equilibrium between droplet and
vapor-gas medium in the time interval between two successive elementary events of emission/absorption of a
molecule by the droplet. 

\par  At present, there exists a complete enough and adequate theoretical description of nonisothermal
nucleation and condensation, both unary$^{1-16}$ and binary,$^{17-20}$ taking into account the above
nonisothermal effects  (especially thoroughly they are investigated in the theory of unary condensation). 
However, as recently pointed out,$^{21}$ there exists an additional nonisothermal effect that may be of
significant importance in  one of the most widespread naturally occurring first-order phase transitions -- the
formation of ubiquitous secondary organic or organic-coated aerosols in the atmosphere via
nucleation/condensation involving atmospheric vapors that are either directly emitted into the atmosphere or
products of gas-phase chemical reactions$^{}$ between both anthropogenic and biogenic organic gaseous
species.$^{22,23}$  

Secondary organic aerosols (SOA)  constitute a large fraction of tropospheric aerosols and  directly
contribute to both scattering and adsorption of solar radiation, having high impact on the Earth climate, air
quality, and human health.$^{24-30}$  The chemical composition of liquid aqueous SOA (only such aerosols are
discussed hereafter) can be extremely  complex,$^{24-27}$
but one can anticipate that the hydrophilic parts of  
organic compounds of an aqueous organic aerosol (OA) particle will be embedded into its aqueous core, 
leaving the hydrophobic parts pointed outward.$^{31,32}$
Surface-located hydrophobic (surfactant) molecules of OA can be processed by their  heterogeneous chemical
reactions with atmospheric gaseous species.$^{33-36}$ The latter may not be directly  involved in
condensation phenomena, but stimulate them by processing hydrophobic patches on the aerosol surface
and rendering it more hygroscopic, thus enabling aerosols to  become cloud condensation nuclei
(CCN);$^{31}$ this is called ``chemical aging" of organic aerosols.  In the atmosphere, the
chemical aging of an OA particle is likely to occur concomitantly$^{21,37,38}$ with the  condensation of water
and other vapors, such as volatile and semi-volatile oxidized organic species. 
These processes depend not only on the composition of the aerosol outer, surface layer, but also on the
physicochemical properties of its core.   

Most of (if not all) heterogeneous chemical reactions on the aerosol surface can be expected to be accompanied
by the release of some enthalpy. The necessity to release such enthalpy of reaction may constitute one of
obstacles hindering a chemical reaction in the gas phase because the surrounding medium is not able to
sufficiently quickly remove the released enthalpy from the reagents/products. The same reaction in the
presence of a third body (aerosol) would occur without such an impediment because a condensed phase particle
would be much more efficient in absorbing the reaction enthalpy. 

Therefore, one can assume that during the chemical aging of a liquid organic aerosol heterogeneous reactions
on its surface are exothermic. Due to the released enthalpy, the   temperature $T$ of a growing aerosol
particle may deviate from
(become higher than)  the ambient (air) temperature $T_a$, i.e., $T>T_a$. 
As recently shown,$^{21}$ under normal atmospheric conditions,  the cooling of the droplet after every such
enthalpy release occurs on timescales longer than the characteristic timescales of droplet evolution with
respect to total number of molecules therein. Consequently, the release of the enthalpy of heterogeneous
reactions involved in chemical aging of organic aerosols leads to the increase of the ability of aerosol
particles to   emit molecules and, hence, causes the decrease of the nucleation rate.  So far, however, this
effect has been barely studied and, consequently, has not been implemented in current atmospheric models. 

Recently, 
we have developed thermodynamic
and kinetic models for the {\em isothermal} formation  of aqueous organic aerosols evolving via both
nucleation/condensation processes and  concomitant chemical reactions on the aerosol surface.$^{37-42}$  
Taking into account the first three reactions in the most probable chemical aging mechanism$^{31}$   
(triggered by atmospheric hydroxyl radicals  abstracting hydrogen atoms from surfactant molecules on the
aerosol surface),    we derived$^{38}$ an explicit  expression for the free energy of formation of a
four-component  aqueous organic aerosol particle  as a function of its  four independent variables of
state.    We also derived a kinetic equation for the size and composition distribution of an ensemble of
aqueous organic droplets, evolving  via nucleation and concomitant chemical aging.$^{41}$ That kinetic
equation explicitly takes account of  chemical reactions on the surface of droplets and hence differs from the
classical kinetic equation of multicomponent  nucleation.  
We showed$^{42}$ that the steady-state solution of this equation subject to appropriate boundary conditions
can be found by using the method of complete separation of variables which was developed in CNT by Kuni et
al.$^{43,44}$ for the kinetics of multidimensional first-order phase transitions {\em without} chemical
reactions.  

In that theory,$^{37-42}$ the temperature of aerosol particles (hereafter referred to as droplets) was assumed
to be constant and equal to the temperature of the surrounding air and, hence, all nonisothermal effects,
involved in concurrent nucleation and chemical aging of organic aerosols  were neglected. In the present work,
we further develop that theory by expanding it to the case where nonisothermal effects are not negligible.

\section{Concurrent nucleation and chemical aging of aqueous organic aerosols}
We will use the formalism of classical nucleation theory (CNT) and treat aerosol particles (droplets) in the
framework  of capillarity approximation,$^{45-47}$   modeling them as spherical  particles of a liquid
multicomponent solution.   Consider an ensemble of such aqueous hydrophilic-hydrophobic organic (AHHO) 
droplets in the air containing three condensable vapors -- water and  hydrophilic and hydrophobic organics
(components 1, 2, and 3, respectively), as well as non-condensable species -- nitrogen oxide,  hydroxyl
radicals, oxygen, and nitrogen dioxide.   Initially, an aerosol contains only  components 1, 2, and 3  due to
their condensation  from the air. The hydrophobic component 3 is mostly  (but possibly not exclusively, if it
also contains a hydrophilic moiety, however weak)  located on the aerosol surface, forming  hydrophobic
patches. However, owing to chemical reactions with atmospheric species,  its molecules   can be transformed
into hydrophilic entities. 

\subsection{Chemical aging mechanism}

According to Ellison et al.$^{31}$, the chemical aging of organic aerosols is most likely
initiated$^{}$   by atmospheric OH radicals abstracting H-atoms from the  hydrophobic moieties of surfactant
molecules on the aerosol surface. (There exist other  pathways of chemical aging,$^{31}$ each 
involving a variety of sequential heterogeneous reactions, but we will not consider them in this work  
because, 
on the one hand, they are less probable and, on the other hand, they can be investigated in 
the same fashion which is presented hereafter). 

Denote a hydrophobic/surfactant molecule by HR, with the radical  ``R\hspace{-0.02cm}{\rad}"  being
the entire  molecule less one of the hydrogen atoms, ``H", in its hydrophobic moiety.
The first three most probable reactions, involved in the chemical mechanism of aerosol aging, are:$^{31}$ 
\beq \mbox{OH (g)+ HR/aerosol}\rightleftharpoons 
\mbox{H$_2$O (g) + R\hspace{-0.02cm}{\rad}/aerosol}.  \eeq
\beq \mbox{O$_2$ (g) + R\hspace{-0.02cm}{\rad}/aerosol}\rightleftharpoons
\mbox{RO$_2$\hspace{-0.16cm}{\rad}/aerosol}. \eeq
\beq 
\mbox{RO$_2$\hspace{-0.16cm}{\rad}/aerosol+NO (g)}
\rightleftharpoons\mbox{RO\hspace{-0.02cm}{\rad}/aerosol+NO$_2$ (g) }. \eeq 

In reaction (1), an OH radical abstracts an H atom from the  hydrophobic moiety of a
surfactant molecule, thus producing a surface-bound radical R\hspace{-0.02cm}{\rad}. 
The latter  is almost immediately oxidized by O$_2$ molecules in reaction (2),  
thus producing a 
surface-bound radical RO$_2$\hspace{-0.16cm}{\rad}. 
The further evolution of radicals
RO$_2$\hspace{-0.16cm}{\rad} may  vary, but always results in the formation of  water soluble and/or volatile
species and/or hydrophilic radicals.$^{}$   Reaction (3) represents one such a pathway  (see ref.31 for the
discussion of various reactive channels of radicals RO$_2$\hspace{-0.16cm}{\rad} and
RO\hspace{-0.01cm}{\rad}).

Reactions (1)-(3) convert a surface hydrophobic molecule HR 
into a radical RO\hspace{-0.02cm}{\rad}. The latter may still contain hydrophobic parts, but there
now appears at least one highly hydrophilic site on its formerly hydrophobic moiety. 
Consequently,  radicals  RO\hspace{-0.02cm}{\rad} will be able to diffuse into the aerosol interior. 
According to numerical evaluations,$^{21}$ the characteristic time of  
sequence (1)-(3) is much shorter than the characteristic time of the evolution of the
total number of molecules in a droplet. Thus, the number of intermediate radicals 
R\hspace{-0.01cm}{\rad} 
(product of reaction (1)) 
and RO$_2$\hspace{-0.1cm}{\rad} 
(product of reaction (2)) 
in
the droplet can be assumed negligible compared to the  number of final radicals RO\hspace{-0.01cm}{\rad}, 
(product of reaction (3)), 
so that sequence  (1)-(3) produces only one additional aerosol component, 
namely, radical  RO\hspace{-0.01cm}{\rad} (component 4). 

Denote the numbers of
molecules of components $1$ (water), $2$ (hydrophilic organic), and $3$ (hydrophobic organic) in a droplet 
by $\nu_1,\nu_2$, and $\nu_3$, respectively, and the number of radicals 
RO\hspace{-0.01cm}{\rad} (component 4) in the droplet  by $\nu_4$. For the sake of simplicity
and uniformity, the radicals RO\hspace{-0.01cm}{\rad} will be also referred to as ``molecules of
component $4$".  

The exothermicity of the gas-phase analogs of reactions (1) and (3) is well-known. 
For example,  in the case where the hydrophobic molecule HR is that of hexanoic acid,  reaction (1) is
accompanied by the release of $\sim 20$ kcal/mol.   Reaction (3) is exothermic with the enthalpy release of
$12$ kcal/mol. The exothermicity of reaction (2) can be conjectured to be
similar to that of reactions (1) and (3). 

Denote the aggregate enthalpy released in a single sequence of reactions (1)-(3) by $\De H_{}$. 
All three reactions have very high reaction rates (either due to a high forward reaction rates and low
backward reaction rates  or high concentration of reagents). 
Thus, for rough evaluations one can assume that the aerosol receives the heat $\De
H_{}$ from the entire sequence of reactions (1)-(3) virtually instantaneously. 

Since the droplet temperature $T$ may now vary (deviate up from the air temperature $T_0$, $T>T_0$) due to
different nonisothermal effects,   it is necessary to introduce a corresponding independent variable of state 
of a droplet. Every droplet will thus have five independent variables of state: the numbers of molecules of
components 1,2,3,4, therein and a temperature related variable.  
We will choose it to be the droplet thermal energy $E$; it
is linear in the  temperature, and will be measured from its value at the temperature $T_0$ of the vapor-gas 
medium. Expressing all the quantities of dimensions of energy in units of $k_BT_0$ ($k_B$ is 
Boltzmann's constant), we have
\beq 
E=c\nu(T/T_0-1),  
\eeq
where 
$\nu=\sum_{i=1}^4\nu_i,\;\; c=\sum_{i=1}^4\chi_ic_i,\;\;\;\chi_i=\nu_i/\nu \;\;(i=1,..,4)$, and  
$c_i\;\;(i=1,..,4)$ is the molecular heat capacity of component $i$ in a droplet (hereafter all heat
capacities are expressed in units of $k_B$). 

\subsection{Free energy of formation of an aqueous organic aerosol via concurrent nucleation and 
chemical aging}

Having chosen $\nu_1,\nu_2,\nu_3, \nu_4$, and $E$ as the independent  variables of state of a single droplet, 
consider a nascent AHHO droplet in the air, 
composed of a 
ternary mixture of condensable vapors -- water and  low-volatility hydrophilic and hydrophobic organics
(components 1, 2, and 3, respectively), as well as non-condensable gases -- nitrogen oxide (component 5), 
hydroxyl radicals (component 6), oxygen (component 7), and nitrogen dioxide (component 8). 
As noticed by Kuni {\em et al.}$^{16}$ and Kurasov,$^{20}$ in the framework of capillarity approximation 
the free energy of nonisotermal formation $F(\nu_1,..,\nu_4,E)$ of such a droplet can be represented as 
(recall that all quantities having the dimensions of energy are expressed in 
units of $k_BT_0$)
\beq F(\nu_1,..,\nu_4,E)=F(\nu_1,..,\nu_4)+E^2/2c\nu,\eeq
where $F(\nu_1,..,\nu_4)$ is the free energy of {\em isothermal} 
formation of a droplet $\nu_1,..,\nu_4$ (i.e., a droplet in internal thermodynamic equilibrium at the same
temperature as the surrounding air temperature $T_0$).  
The term 
$E^2/2c\nu$ on the RHS of
eq.(5) represents the contribution to $F(\nu_1,..,\nu_4,E)$ due to the deviation of the droplet temperature
$T$ from the air temperature $T_0$ (i.e., it represents the work of heating/cooling the droplet from
temperature $T_0$ to temperature $T$).

An analytic expression for the ``isothermal" free energy $F(\nu_1,..,\nu_4)$ 
was obtained in ref.$^{}$ (see also ref.);  
it can be written as
\beq F\equiv F(\nu_1,\nu_2,\nu_3,\nu_4)=-\sum_{i=1}^{4}\nu_i
\ln\frac{\zeta_i}{\chi_if_i(\chi_1,\chi_2,\chi_3)} 
+\sigma(\chi_1,\chi_2,\chi_3)\;A(\nu_1,\nu_2,\nu_3,\nu_4)/k_BT, \eeq
where 
$\zeta_i=P_i/P_{ie}\;\;(i=1,2,3)$ is the 
saturation ratio of the condensable component  $i$ in air, with $P_i$ being its partial pressure  
and $P_{ie}$ its  equilibrium vapor pressure; 
$f_i(\chi_1,\chi_2,\chi_3)$ is the activity coefficient of component $i$ in the four-component solution of
composition $\chi_1,\chi_2$, and $\chi_3$ 
(mole fractions $\chi_i\;\;(i=1,..,4)$ are related by $\chi_1+\chi_2+\chi_3+\chi_4=1$);  
$\sigma(\chi_1,\chi_2,\chi_3)$ and  $A(\nu_1,\nu_2,\nu_3,\nu_4)$  are the surface tension and surface area of
the droplet of radius $R$. 
The quantity $\zeta_4$ 
can be loosely (for the sake of convenience) called  ``the saturation ratio" of component 4; it  
is defined as
$\zeta_4=(\zeta_3\zeta_5\zeta_6\zeta_7K_{\mbox{\tiny{eq}}})/(\zeta_1\zeta_8)$,  where $K_{\mbox{\tiny{eq}}}$
is the aggregate equilibrium constant of sequence (1)-(3), 
and  $\zeta_j=P_j/P_{j0}\;\;(j=5,6,7,8)$, with
$P_j$ the partial pressure of component $j$ in the air and  $P_{j0}$ its standard partial pressure  for
which the standard  Gibbs free energy change (at temperature $T_0$) 
of reactions (1)-(3) is assumed to be known. 

The function  $\widetilde{F}=\widetilde{F}(\nu_1,\nu_2,\nu_3,\nu_4,E)$ determines a free-energy surface in a
6-dimensional  space.  Under conditions when a multicomponent first-order phase transition occurs via {\em
homogeneous  nucleation} (as assumed hereinafter), it  has a shape of a hyperbolic paraboloid (``saddle-like"
shape  in three dimensions).  Hereafter, the extremum of this surface will be referred to as the ``saddle
point" and all quantities at this point  will be marked with the subscript ``c".  An aerosol particle whereof
the variables $(\nu_1,\nu_2,\nu_3,\nu_4,E)$ coincide with the respective  coordinates of the saddle point is
referred to as a ``nucleus". The nucleus characteristics  $\nu_{1c},\nu_{2c},\nu_{3c},\nu_{4c},E_c$ are thus
determined as the solution of five simultaneous equations 
\beq 
\left. \widetilde{F}'_i(\{{\bf \nu}\},E) \right|_c=0\;\;\; (i=1,..,4),\;\;\; 
\left. \widetilde{F}'_{\mbox{E}}(\{{\bf \nu}\},E)\right|_c=0, 
\eeq 
where $\widetilde{F}'_i(\{{\bf \nu}\},E)=\left. \partial \widetilde{F}(\{{\bf \nu}\},E)/\partial
\nu_i\right|_{\widetilde{\nu_i}=const,E=const},\;\; \widetilde{F}'_E(\{{\bf \nu}\},E)=
\left.\partial \widetilde{F}(\{{\bf \nu}\},E)/\partial E \right|_{\{{\bf \nu}\}=const}$, 
and  we introduced the notations $\{{\bf \nu}\}$ for the
set of variables $\nu_1,\nu_2,\nu_3,\nu_4$ (such that $\{{\bf \nu}\}\equiv (\nu_1,\nu_2,\nu_3,\nu_4)$) and
defined $\widetilde{ \nu_i}$ as a composite variable obtained by excluding $ \nu_i$ from $\{{\bf \nu}\}$, so
that $\{{\bf \nu}\}=( \widetilde{\nu_i},\nu_i)=(\nu_1,\nu_2,\nu_3,\nu_4)$. As clear from eqs.(5) and (6),
$E_c=0$.   Note that eqs.(5) and (6) involve 
only approximations intrinsic to the capillarity approximation.$^{45,46}$

During nucleation, aerosol particles overcome a free-energy barrier (6D surface 
determined by the function $\widetilde{F}=\widetilde{F}(\{\nu\},E)$) 
to become irreversibly growing droplets. 
The crucial role in this process is played by the evolution of aerosol particles in the saddle-point region
of the space of variables $\nu_1,..,\nu_4,E$.$^{}$  At this stage droplets are assumed to be large
enough to be treated in the capillarity approximation, conventionally used in the framework of CNT.
In the isothermal CNT, the saddle-point region itself is defined as the vicinity of the saddle point in which
the bilinear approximation for $F(\{\nu\})$ is
acceptable,$^{46,48-53}$ 
\beq F(\{\nu\})=F_c+\frac1{2}\sum_{i,j=1}^4F''_{ijc}(\nu_i-\nu_{ic})(\nu_j-\nu_{jc}), \eeq 
where $F\equiv F(\{{\bf \nu}\})$, $F''_{ij}=\partial^2 F/\partial\nu_i \partial\nu_j\;\;(i,j=1,..,4)$. 
In this approximation, the first derivatives
$F'_i\equiv F'_i(\{\nu\})=\partial F/\partial\nu_i \;\;(i=1,..,4)$  are linear superpositions of 
deviations $(\nu_i-\nu_{ic}) \;\;(i=1,..,4)$.  

\subsection{Temporal evolution of an ensemble of aqueous organic droplets with variable temperature} 
Consider an ensemble of AHHO droplets (evolving
via both nucleation and concomitant chemical aging) 
and  denote their distribution function with   respect to $\nu_1,\nu_2,\nu_3,\nu_4,E$  
at time $t$ by $g(\nu_1,\nu_2,\nu_3,\nu_4,E,t)$.  According to the definition of $\{{\bf \nu}\}$ 
and  $\widetilde{\nu}_i$  
any function $f$ of variables  $\nu_1,\nu_2,\nu_3,\nu_4$ may be denoted as either 
$f(\nu_1,\nu_2,\nu_3,\nu_4)$ or  $f(\{\nu\})$ or $f(\widetilde{\nu}_i,\nu_i)$; 
e.g., $g(\{{\bf \nu}\},E, t)\equiv
g(\widetilde{\nu}_i,\nu_i,E,t)\equiv g(\nu_1,\nu_2,\nu_3,\nu_4,E,t)$.

A differential equation, governing the temporal evolution of an ensemble of such  aqueous organic droplets 
(with the distribution function $g(\{{\bf \nu}\},E, t)$) and taking into account the relevant nonisothermal
effects, can be derived by combining the procedure   used to derive the kinetic equation of the isothermal
process$^{41,42}$ with the procedure of  Kuni and Grinin$^{5-10,54}$ for the derivation of the kinetic equation
of nonisothermal unary nucleation. First, it is necessary to construct a discrete five-dimensional balance
equation of nonisothermal ternary nucleation and concomitant chemical aging  taking into account all types of
elementary interactions of nascent droplets   with the vapor-gas medium (air). 

\subsubsection{Discrete equation of balance for the distribution function}

As usual in the simplest version of CNT, let us assume that the metastability of the vapor mixture is
created instantaneously and does not change during the whole nucleation--chemical aging process process.
The temperature $T_0$ of the vapor-gas medium (air) and the number density of passive
gas molecules are also fixed. At the nucleation stage, the droplets are so small that 
the timescale of their internal relaxation processes are very small in comparison with 
the timescale between successive elementary interactions between droplet and air and even smaller compared to 
the timescale between two successive sequences of chemical reactions (1)-(3);  
elementary interactions between droplets and air are assumed to take place under 
a free-molecular regime. This allows one to assume that a liquid droplet attains its internal 
thermodynamical equilibrium  before each successive interaction with the vapor-gas medium and before each
sequence of reactions (1)-(3).  

Let $\eps$ be the thermal energy of molecules striking a nucleus and let $\eps'$ be the thermal energy of
molecules reflected or emitted  by a nucleus. Since the times of internal relaxation processes of nucleus are
small, the number $W_i^-\;\;(i=1,2,3)$ of molecules of component $i$ which a droplet emits per unit time as
well as the distribution $w'$ of the emitted or reflected molecules with respect to their energy $\eps'$ are
determined (assuming the complete thermal adaptation of reflected molecules) by the droplet energy: 
$W_i^-\equiv W_i^-(\{\nu\},E)\;\;(i=1,2,3),\;\;\;w'\equiv
w'(\{\nu\},E|\eps')$.
Here the variables $\nu_i\;\;(i=1,..,4)$, and $E$ correspond to the state of the nucleus  before  the
interaction (because the temperature fluctuation  effect and the effect of nucleus
thermal quasi-isolation compensate each other$^{6}$). 
On the other hand, 
the distribution $w\equiv w(\eps)$ of molecules 
striking a droplet with respect to their energy $\eps$, and 
the number $W_i^+\equiv W^+_i(\{{\bf \nu}\})\;\;(i=1,2,3)$ of molecules of component $i$ that  the  droplet
absorbs from  air per unit time are independent of the droplet temperature; they are both determined by the
temperature $T_0$ of the vapor-gas medium. 

The material and thermal exchange between droplet and air occurs via the following elementary interactions:\\ 
(a123) absorption of a molecule of component $1$ or $2$ or $3$ from the air into the droplet  $\{\nu\}$ with
the rate $W^+_i\;\;(i=1,2,3)$ accompanied by the release of the latent heat of
condensation  $\beta_i\;\;(i=1,2,3)$ to the droplet (recall that all quantities having the
dimension of energy are expressed in units of $k_BT_0$);\\  
(e123) emission of a molecule of component $1$ or
$2$ or $3$ from the aerosol $\{\nu\}$  into the air with the rate $W^-_i\;\;(i=1,2,3)$, accompanied by the
removal of the latent heat of evaporation/condensation  $\beta_i\;\;\;(i=1,2,3)$ from the droplet;\\  
(f4) production of a
``molecule" of component 4 (radical RO\hspace{-0.01cm}{\rad}) via the  forward sequence of heterogeneous
chemical reactions (1)-(3) on the surface of aerosol  $\{\nu\}$, with the rate
$W^+_4=W^+_4(\{\nu\},E)$, accompanied by the
release of the aggregate enthalpy of reactions (1)-(3)  $\beta_4\equiv \De H_{}/k_BT_0$ from the droplet;\\ 
(b4) destruction of a ``molecule" of component 4 (radical RO\hspace{-0.01cm}{\rad})
via the  backward sequence of chemical reactions (1)-(3) on the surface of droplet  $\{\nu\}$,
with the rate $W^-_4=W^-_4(\{\nu\},E)$, accompanied by the
removal of the aggregate enthalpy of reactions (1)-(3)  $\beta_4$ from the droplet;\\  
(r123g) reflection of a molecule of the vapor-gas medium (components 1, 2, 3, and passive gas). 

Thus, one can write 
the initial discrete equation of balance, governing the evolution of the distribution 
$g(\{{\bf \nu}\},E,t)$, as 
\beq \frac{\partial g(\{{\bf \nu}\},E,t)}{\partial t}=\sum_{i=1}^3D'_i+D_4 + D_{\mbox{\tiny E}},  \eeq
where
\be
D'_i=\int_0^\infty\,d\eps\; [W_i^+(\widetilde{\nu_i},\nu_i-1)w(\eps)g(\widetilde{\nu_i},\nu_i-1,E-\beta_i-
\eps) -
W_i^+(\{{\bf \nu}\})w(\eps)g(\{{\bf \nu}\},E,t)] \nonumber \\
+\int_0^\infty\,d\eps'\;
[W_i^-(\widetilde{\nu_i},\nu_i+1,E+\beta_i+\eps')w'(\widetilde{\nu_i},\nu_i+1,E+\beta_i+\eps'|\eps')
\times
\nonumber \\  g(\widetilde{\nu_i},\nu_i+1,E+\beta_i+\eps',t)
 -W_i^-(\{{\bf \nu}\},E)w'(\{{\bf \nu}\},E|\eps')
g(\{{\bf \nu}\},E,t)]\;\;\;\;(i=1,2,3), 
\ee
\be 
D_4= W^+_4(\nu_1,\nu_2,\nu_3+1,\nu_4-1,E-\beta_4)g(\nu_1,\nu_2,\nu_3+1,\nu_4-1,E-\beta_4,t)- 
W^+_4(\{{\bf \nu}\},E)g(\{{\bf \nu}\},E,t) \nonumber \\
+  
W_4^-(\nu_1,\nu_2,\nu_3-1,\nu_4+1,E+\beta_4)g(\nu_1,\nu_2,\nu_3-1,\nu_4+1,E+\beta_4,t)- 
W_4^-(\{{\bf \nu}\},E)g(\{{\bf \nu}\},E,t), 
\ee 
\beq  D_{\mbox{\tiny E}}=\int_0^\infty\int_0^\infty\,d\eps'd\eps
W^{\mbox{\tiny ref}}_{}(\{{\bf \nu}\})[w(\eps)w'(\{{\bf \nu}\},E-\eps+\eps'|\eps')
g(\{{\bf \nu}\},E-\eps+\eps')
-w(\eps)w'(\{{\bf \nu}\},E|\eps')g(\{{\bf \nu}\},E)],\eeq
\beq
W_{\mbox{\tiny rfl}}(\{{\bf \nu}\})=\sum_{i=1}^3\frac{1-\alpha_{ci}}{\alpha_{ci}}\alpha_{ti}
W_i^+(\{{\bf \nu}\})+\alpha_{g}W_g(\{{\bf \nu}\}) \;\;, \eeq
$\alpha_{ci}$ and $\alpha_{ti}\;(i=1,2,3)$ are the condensation  (sticking) coefficient and coefficient of
thermal adaptation  in a reflection event, respectively, of  molecules of component $i$; $\alpha_{tg}$ is the
coefficient of thermal adaptation  in a reflection event of a passive gas molecule;  $W_g(\{{\bf \nu}\})$ is
the number of molecules of the passive gas impinging on a droplet per unit time.  Clearly, $W_{\mbox{\tiny
rfl}}$ would determine the total number of molecules reflected by the droplet per unit time if
$\alpha_{ti}\;\;(i=1,2,3)$ and  $\alpha_g$ were all equal to $1$. Note again that  these equations assume the
evolution of aerosols to occur through the absorption from and emission into the vapor of single molecules of
components $1,2$,  and $3$  (i.e., multimer absorption  and emission are neglected),  as well as through the
single sequences (1)-(3) of forward and backward reactions whereby a radical  RO\hspace{-0.01cm}{\rad} is
either formed or destroyed. 

The terms $D'_1,D'_2$, and $D'_3$ on the RHS of eq.(9) represent the contributions to  $\partial
g(\{\nu\},E,t)/\partial t$  from the material exchange events of type (a123) and (b123), whereas the term
$D_4$  represents the contributions to $\partial g(\{\nu\},E,t)/\partial t$ from the elementary events of type
(f4) and (b4); the term $D_E$ reresents the contribution to $\partial g(\{\nu\},E,t)/\partial t$ from
elementary events of type (r), i.e., from the the kinetic/internal energy exchange between droplets and all 
molecules reflected from droplets without being absorbed by them.  Furthermore, on the RHS of the each of
eq.(8)  the first integral represents the contributions to $\partial g(\{\nu\},t)/\partial t$ from the
absorption events (a123),  whereas the second integral is due to the emission of molecules from aerosols into
the air. On the RHS of eq.(11), the first two terms represent the contributions to $\partial
g(\{\nu\},t)/\partial t$ from the forward sequences (1)-(3) of chemical reactions on aerosols (whereby
radicals RO\hspace{-0.01cm}{\rad} are produced), whereas the third and fourth terms therein are due to the
backward sequences (1)-(3) (whereby radicals RO\hspace{-0.01cm}{\rad} are destroyed). As clear from eq.(11)
and in consistency with the sequence of chemical reactions (1)-(3), the change of the aerosol distribution due
to the variable $\nu_4$ is always accompanied by its change due to the variable $\nu_3$, while the latter can
also change independently due to the direct material exchange between aerosols and air. 

\par Expanding eqs.(10)-(12) in Taylor series in the deviations of nucleus
characteristics from $\nu_1,\nu_2,\nu_3,\nu_4$, and $E$, after simple transformations  one can reduce eq.(9) 
to 
\beq \frac{\partial g(\{{\bf \nu}\},E,t)}{\partial t}=\sum_{i=1}^4D_i+\frac{\partial}{\partial
E}I_E, \eeq
where 
\be
D_i=W_i^+(\widetilde{\nu_i},\nu_i-1)g(\widetilde{\nu_i},\nu_i-1,E-\beta_i) -
W_i^+(\{{\bf \nu}\})g(\{{\bf \nu}\},E,t) \nonumber \\
+W_i^-(\widetilde{\nu_i},\nu_i+1,E+\beta_i) g(\widetilde{\nu_i},\nu_i+1,E+\beta_i,t)
 -W_i^-(\{{\bf \nu}\},E)
g(\{{\bf \nu}\},E,t)\;\;\;\;(i=1,2,3), 
\ee 
\beq I_E=	
-\sum_{i=1}^{3}\frac{\beta_i^2}{k_i}W_i^+\left(\frac{E}
{c\nu}+\frac{\partial}{\partial E}\right)g(\nu_1,\nu_2,E),\eeq
\beq
\frac1{k_i}=\frac{\tilde{c}_i}{\alpha_{ci}\beta_i^2}[\alpha_{ci}+
\alpha_{ti}(1- 
\alpha_{ci})+\alpha_{tg}p_i]\;\;(i=1,2,3), \eeq
\beq
p_i=\left(\frac{m_i}{m_g}\right)^{1/2}\frac{n_gc_g}{2n_i\widetilde{c}_i}\;\;(i=1,2,3),\eeq 
$\widetilde{c}_i,
m_i$, and $n_i \;(i=1,2,3)$ are the effective (in the  sense of energy transfer to the nucleus) heat capacity,
mass, and number density of molecules, respectively, of component $i$ of the vapor mixture; $c_g, m_g$, and
$n_g$ are the analogous values of the passive gas.

The terms $D_1,D_2$, and $D_3$ in eq.(14) describe the simultaneous transfer of both the substance and the
latent heat of condensation to droplets by the molecules  of condensable components of the air ($1,2$, and
$3$, respectively).  These terms have the structure characteristic of the Zeldovich--Frenkel nucleation
equation, but modified to take into account that in every adsorption or emission of a molecule of component
$i\;\;(i=1,2,3)$ by a droplet not only does the corresponding $\nu_i$ change by $\pm 1$, but also the variable
$E$ changes by $\pm\beta_i$. Having a similar structure, the term $D_4$ in eq.(14)  takes into account that in
every sequence of reactions (1)-(3) the change of the variable $\nu_4$ by $\pm 1$ is accompanied by changes in
$\nu_2$ and $E$ by $\mp 1$ and $\pm \beta_4$, respectively.

The term $-\partial I_E/\partial E$ in eq.(14) describes the transfer of  the kinetic and internal energies to
the droplets by all the molecules of the vapor-gas medium. Its Fokker-Planck form corresponds to the
fulfillment  of the condition 
\beq 1/(c\nu)^{1/2}\ll 1, \eeq 
meaning the smallness of the energy transfer by a molecule of the vapor-gas medium in comparison with the rms
fluctuation of the droplet energy  (recall that, according to the thermodynamic theory of fluctuations,  
$(c\nu)^{1/2}=(\sum_{i=1}^4c_i\nu_i)^{1/2}$ represents the rms fluctuation of the droplet energy in the
absence of material exchange between droplet and vapor mixture). 

\subsubsection{Kinetic equation of {\em nonisothermal} concurrent nucleation and chemical aging}
\par Let us introduce the variable $\xi$ instead of variable $E$ as 
\beq \xi=\frac{E}{(2c\nu)^{1/2}} \eeq
and present the distribution $g(\{{\bf \nu}\},E)$ in the form
\beq g(\{{\bf \nu}\},E,t)=[2\pi
c\nu]^{-1/2}e^{-\xi^2}P(\{{\bf \nu}\},\xi,t), \eeq
where the function $P(\{{\bf \nu}\},\xi,t)$ of $\{{\bf \nu}\},\xi,t$ will be referred to as ``the 
distribution of droplets with respect to $\{{\bf \nu}\}$ and $\xi$ at time $t$", although, strictly speaking, such a
distribution is represented by the product $\pi^{-1/2}\mbox{e}^{-\xi^2}P(\{{\bf \nu}\},\xi,t)$.

\par Usually $\beta_i\gg 1\;(i=1,..,4)$, so the parameter
\beq \alpha_i=\frac{\beta_i}{(2c\nu)^{1/2}}\,\,\;\;(i=1,..,4) \eeq
will not be small despite inequality (19). For the three condensable air components  
$\alpha_j\;\;(j=1,2,3)$ represents 
the relative latent heat of condensation/vaporization of component $j$ per molecule, i.e. the 
latent heat of component $j$ per molecule expressed in units of $(2c\nu)^{1/2}$, rms 
fluctuation of a droplet energy (in the
absence of material exchange between droplet and vapor mixture) multiplied by $\sqrt{2}$; likewise, for component 4 (product of the sequence of reactions (1)-(3)), 
$\alpha_4$ represents 
the relative aggregate enthalpy released in a single sequence of reactions (1)-(3), i.e. the 
aggregate enthalpy of reactions (1)-(3) expressed in units of $(2c\nu)^{1/2}$.
Although $\alpha_i\;\;(i=1,..,4)$ is always smaller than $1$, in order of magnitude $\alpha_i\sim 1$ 
(all $\alpha_i\;\;(i=1,..,4)$ are assumed to be constant and equal to their 
values for the nucleus$^{5,16}$). 

\par Equation (14) can be transformed into a differential equation for the distribution 
$P(\{{\bf \nu}\},\xi,t)$ in a standard way,$^{5,8}$ by expanding the terms $D_i\;\;(i=1,..,4)$ therein 
in Taylor series 
in the deviations of $\nu_i\pm 1$ from $\nu_i$ and $E\pm \beta_i$ from $E$ 
$(i=1,..,4)$ and (as usual in CNT) assuming that with respect to the variables 
$\nu_i\;\;(i=1,..,4)$ the resulting differential equation has the form of the Fokker-Planck equation 
(with linear force coefficients)
(see ref.53 for more detais):
\newpage 
\be \frac{\partial P}{\partial t} &=& \\
&&-\sum_{i=1}^4\frac{\partial }{\partial \nu_i}\left(\hat{L}_i -
W_i^+ \sum_{m=1}^\infty\,\frac{\alpha_i^m}{m!}\frac{\partial^m}{\partial
\xi^m}\right)P  \;\;\hspace{8cm} (a)\nonumber \\ 
&&+\left(\frac{\partial }{\partial \nu_3}\hat{L}_{43}^{}-\frac{\partial }{\partial \nu_3}\hat{L}_{4}^{} 
-\frac{\partial }{\partial \nu_4}\hat{L}_{43}^{}\right)\;P\;\;\hspace{8.3cm}(b)\nonumber\\
&&-\frac{\partial }{\partial \nu_3}W_4^+ \sum_{m=1}^\infty\,\frac{\alpha_i^m}{m!}\frac{\partial^m}{\partial
\xi^m}P\;\;\hspace{10.5cm}(c)\nonumber\\
&&+ \sum_{i=1}^4\sum_{m=1}^\infty\;\frac{(-1)^m\alpha_i^m}{m!}\hat{L}_i
\left(\frac{\partial}{\partial\xi}-2\xi\right)^mP \;\;\hspace{8cm}(d) \nonumber \\
&&+ \sum_{m=1}^\infty\;\frac{(-1)^m\alpha_4^m}{m!}\hat{L}_{43}^{}
\left(\frac{\partial}{\partial\xi}-2\xi\right)^mP \;\;\hspace{8.5cm}(e) \nonumber \\
&&- \sum_{i=1}^4W_i^+\sum_{m\ne m'=1}^{\infty} \frac{(-1)^m\alpha_i^{m+m'}}{m!m'!}
\left(\frac{\partial}{\partial\xi}-2\xi\right)^{m'}\frac{\partial^m }{\partial \xi^m}P \;\;\hspace{6cm}(f) 
\nonumber \\
&&+\left[\left(\sum_{i=1}^3\frac{k_i+1}{k_i}W_i^+\alpha_i^2+
W_4^+\alpha_4^2\right)\left(\frac{\partial}{\partial\xi}-2\xi\right)\right.
\frac{\partial}{\partial\xi} \;\;\hspace{6.5cm} (g1) \nonumber\\
&&-\left.\sum_{i=1}^4\sum_{m=2}^\infty\;W_i^+\frac{(-1)^m\alpha_i^{2m}}{m!m!} 
\left(\frac{\partial}{\partial\xi}-2\xi\right)^m
\frac{\partial^m}{\partial\xi^m}\right]P\;\;\hspace{6.5cm} (g2) \nonumber 
\ee 
(for the sake of simplicity of notation,  the independent arguments $\nu_1,..,\nu_4,\xi$, and $t$ of 
$W_i^+\;\;(i=1,..,4)$ and $P$ are omitted),   
where  we introduced the operators
\beq 
\hat{L}_{43}^{}\equiv -W_4^+(F_3'+\frac{\partial}{\partial \nu_3}),\;\; 
\hat{L}_i\equiv -W_i^+(F_i'+\frac{\partial}{\partial \nu_i})\;\;\;\;(i=1,..,4), 
\eeq
with $F_i'\equiv \partial F/\partial
\nu_i$, and $F$ the free energy of formation of a droplet with 
characteristics $\nu_1,\nu_2,\nu_3,\nu_4$, and $E=0$: $F\equiv F(\{{\bf \nu}\},E=0)$). 

Hereafter, we will be interested only in the saddle point region  
$|\nu_i-\nu_{ic}|\lesssim \Delta\nu_{ic}$, because the evolution of droplets there 
plays the determining role for the nucleation 
kinetics.$^{5-12,43,44,46,48-52}$ The half-width of this region $\Delta\nu_{ic}\;\;(i=1,..,4)$ represents a 
characteristic scale of change of $\nu_i$-dependent functions and, consequently, 
we have the operator estimate ${\partial}/{\partial \nu_i}\sim 
1/{\Delta\nu_{ic}}$. On the other hand, $|F_i'|\lesssim 
1/\Delta\nu_{ic}\;(i=1,..,4)$ in this region, according to eq.(8). Therefore, in this   
region, the second term on the RHS of eq.(27) is dominant, which substantiates the following operator
estimates:
\beq \frac1{W_i}\hat{L_i} \sim \frac{\partial}{\partial \nu_i} \sim \frac1{\Delta\nu_{ic}}\;\;
\;\;(i=1,..,4). \eeq
The terms $\partial L_iP/\partial \nu_i\;(i=1,..,4)$, in the RHS of eq.(27) have the second
order of smallness in $1/\Delta\nu_{ic}$, but they are retained in the framework of CNT because they are
necessary for the self-consistent description of the kinetics of nucleation. 

According to eq.(21), the characteristic values of
$\xi$ lie in the interval $|\xi|\lesssim 1$, where we have the estimates   
\beq  \partial /\partial \xi \sim \xi \sim 1. \eeq
Therefore, $\alpha_i\;\;(i=1,..,4)$ are the expansion parameters in the series  in $m$ and $l$.  Retaining all
the terms of those series means that we extend  the theory to values $\alpha_i\sim 1\;\;(i=1,..,4)$ and hence
go beyond the framework of the Fokker-Planck approximation.

Let us establish a relative importance of the terms on the RHS of eq.(23), taking into account the estimates
(25),(26), and  $|F_i'|\lesssim 1/\Delta\nu_{ic}\; (i=1,..,4)$.  Since the terms (a)-(e) on the RHS of
eq.(23)  contain the operators $L_i,\partial /\partial \nu_i \;\;(i=1,..,4), L_3^{(4)}, L_4^{(4)}$,  we
conclude that their ratios to the  last, seventh, term do not exceed  $1/\Delta\nu_{ic}\;\;(i=1,..,4)$  (which
are much smaller than $1$) in order of magnitudes. 

Comparing the first member of the last term (g) on the RHS of eq.(23) with the second member therein, we
conclude that the first member is the main one in this term, because of the inequalities $(k_i
+1)/k_i>1\;\;(i=1,..,4)$ and factorials $m!m!$. One can also see that the ratio of the term (f) to the
last term (g) does not  exceed the parameter
$$  
\delta_{\tiny g}^{\tiny f} \equiv \frac{\sum_{i=1}^4W_i^+\alpha_i^3}{2\left(\sum_{i=1}^3W_i^+\alpha_i^2
(k_i+1)/k_i+W_4^+\alpha_4^2\right)}
$$
Assuming it to be much smaller than unity, $ \delta_{\tiny g}^{\tiny f} \ll 1$, 
one can conclude that on the RHS of eq.(23) the last term is the predominant one. 

Equation (23) governs the time evolution of the five-dimensional distribution
$P$. The hierarchy of terms established above corresponds to the
hierarchy of time scales in the  development of this distribution.

Denote the principal operator of the governing equation (23), i.e., 
the operator of the dominant term on its RHS, by $\hat{\Lambda}$:
\beq 
\hat{\Lambda}=\left(\sum_{i=1}^3\frac{k_i+1}{k_i}W_i^+\alpha_i^2+
W_4^+\alpha_4^2\right)\left(\frac{\partial}{\partial\xi}-2\xi\right) 
\frac{\partial}{\partial\xi} 
-\sum_{i=1}^4\sum_{m=2}^\infty\;W_i^+\frac{(-1)^m\alpha_i^{2m}}{m!m!} 
\left(\frac{\partial}{\partial\xi}-2\xi\right)^m
\frac{\partial^m}{\partial\xi^m}.  
\eeq
One can see that the eigenfunctions of this operator are the Hermite polynomials 
$H_j\equiv H_j(\xi)$ ($H_0=1,\;H_1=2\xi,\; H_2=4\xi^2-2, ...$), satisfying the recursion relations 
\beq \frac{\partial}{\partial \xi}H_j=2jH_{j-1},\, \left(\frac{\partial}{\partial
\xi}-2\xi\right)H_j=-H_{j+1},\;\;\;(j=1,2,...) \eeq
so that  
\beq 
\hat{\Lambda}H_j=\Lambda_jH_j\;\;\;(j=0,1,...), 
\eeq 
where $\Lambda_j\;\;(j=0,1,2,...)$ is the eigenvalue, corresponding to the eigenvector $H_j$. 
As clear from eqs.(27) and (28),
\beq 
\Lambda_j=-j\lambda_j,\;\;\;
\lambda_j=
2\left(\sum_{i=1}^3\frac{k_i+1}{k_i}W_i^+\alpha_i^2+
W_4^+\alpha_4^2\right) +(j-1)!\sum_{i=1}^4\sum_{m=2}^\infty\;W_i^+\frac{(2\alpha_i^{2})^m}{m!m!(j-m)!} 
\eeq 
(for $j=0$ and $j=1$ the sum over $m$ on the RHS of the latter equality is absent; $0!=1$ is adopted). 
Since $0<\lambda_0<\lambda_1<\lambda_2<...$, one can conclude that all the eigenvalues 
$\Lambda_j$ with $j=1,2,...$ are negative, 
whereas the eigenvalue $\Lambda_0$ is equal to zero: $\Lambda_0=0,\;\;\Lambda_j<0\;\;(j=1,2,...)$.

The Hermite polynomials form a complete system of eigenfunctions (an orthogonal basis)  
satisfying the orthogonality and normalization relations
\beq (H_j,H_k)=\delta_{jk}2^jj!\;\;\;(j,k=0,1,2,...), \eeq
where $\delta_{jk}$ is the Kronecker delta and the scalar product $(\Phi,\Psi)$ of
function $\Phi$ and $\Psi$ of $\xi$ is defined as 
\beq (\Phi,\Psi)=\pi^{-1/2}\int_{-\infty}^\infty\;d\xi\,e^{-\xi^2}\Phi\Psi. \eeq

As follows from eqs.(21),(32), and $H_0=1$, the four-dimensional distribution
$f\equiv f(\nu_1,...,\nu_4)$ of droplets with respect to variables $\nu_1,..,\nu_4$ is given by the equation
$f=(H_0,P)$, i.e. the four-dimensional distribution $f$ is the projection of the five-dimensional distribution 
$P$ on $H_0$. Taking this into account, let us take the projection of governing equation (23) on $H_0$.
According to eqs.(28) and (31), non-zero contributions to this projection arise only from 
the first member of the first term, (a), of the order 
of $1/(\Delta\nu_{ic})^2\;\;(i=1,..,4)$. One can thus obtain 
\beq \frac{\partial f}{\partial t}= - \sum_{i=1}^4\frac{\partial J_i}{\partial \nu_i}, \eeq
where
\beq 
J_i=\left(H_0, \left(\hat{L}_i -
W_i^+ \sum_{m=1}^\infty\,\frac{\alpha_i^m}{m!}\frac{\partial^m}{\partial
\xi^m}\right)P\right)\;\;\;\;(i=1,2),
\eeq 
\beq 
J_3=\left(H_0, \left[\left(\hat{L}_3 -
W_3^+ \sum_{m=1}^\infty\,\frac{\alpha_3^m}{m!}\frac{\partial^m}{\partial \xi^m}\right)
+\left(\hat{L}_{43}^{}-\hat{L}_4^{}+W_4^+\sum_{m=1}^\infty\,\frac{\alpha_3^m}{m!}
\frac{\partial^m}{\partial \xi^m}\right)\right]P\right), 
\eeq  
\beq 
J_4=\left(H_0, \left[\left(\hat{L}_4 - 
W_4^+ \sum_{m=1}^\infty\,\frac{\alpha_4^m}{m!}\frac{\partial^m}{\partial  \xi^m}\right) +  
(-\hat{L}_{43}^{})\right]P\right), 
\eeq 
is the (averaged over $\xi$) flux  of nuclei along the $\nu_i$-axis.

\subsubsection{The stage of thermal relaxation}

\par Retaining on the RHS of governing equation (21) only the leading term (g) (containing two members), and
taking into account definition (21), we obtain 
\beq
\frac{\partial P}{\partial t}=\hat{\Lambda}P\
\eeq
The solution of this equation is given, according to relation (29), by 
\beq P=f+ \sum_{j=1}^\infty e^{-j\lambda_jt}f_jH_j,\eeq
where $f$ and $f_j$ are independent of $\xi$ and $t$ and can be
presented, by virtue of eq.(32), as
\beq f=(H_0,P)=(H_0,P|_{t=0}) , f_j=(2^jj!)^{-1}(H_j,P|_{t=0}) \eeq 
($P|_{t=0}$ is the three dimensional distribution $P$ at $t=0$).
From $f=(H_0,P)$ and eq.(43) it follows that $f$ still represents the four-dimensional 
distribution of droplets with respect to $\nu_1,..,\nu_4$ and it does not 
change as long as $P$ is governed by eq.(41).
Therefore, eq.(42) describes the stage  of thermal relaxation of droplets; this stage is
characterized by the spectrum of relaxation times $1/j\lambda_j\;\;(j=1,2,...)$, which decrease
with increasing $j$.

As follows from eqs.(21), (32) and $f=(H_0,P)$, if $\Phi$ is some function of the variable $\xi$, its average
value $\Phi$ with respect to the variable $\xi$ is determined as $ \bar{\Phi}=(\Phi,P)/f$. Therefore, average
values with respect to $\xi$ will also change  together with $P$ in the process of thermal relaxation.

According to eq.(38), at the end of the thermal relaxation 
\beq P\simeq f \;\;\;(t\gtrsim t_{\xi}), \eeq
\beq t_{\xi}=1/\lambda_1=
\frac1{2}\left(\sum_{i=1}^3\frac{k_i+1}{k_i}W_i^+\alpha_i^2+
W_4^+\alpha_4^2\right)^{-1}, 
\eeq
where $t_{\xi}$ is the principal thermal relaxation time. Since $f$ does not depend on $\xi$, by virtue of 
eqs.(21),(40), we can conclude that the distribution of droplets with respect to temperature approaches a
quasiequilibrium Gaussian distribution by the end of the thermal relaxation stage, whereof the duration  is
given by $t_{\xi}$. The inverse quantity $1/t_{\xi}$ determines the ``speed" of thermal relaxation, and it
contains the contributions from the latent heat of condensation and the enthalpy of chemical reactions, as
well as from the exchange of kinetic (thermal) energy between droplets and molecules of vapor-gas medium.

Denote by $t_{\nu}$ the characteristic time of change of the size (four-dimensional)  distribution $f$. In
order to obtain an estimate for $t_{\nu}$ at the end of  the stage of thermal relaxation, let us replace $P$
by $f$ in eqs.(34)-(36) (which is an accurate   enough approximation by virtue of eq.(40)) and then substitute
$J_i\;\;(i=1,..,4)$ in  eq.(33). Using estimates (25), one can obtain for  $t_{\nu}$: 
\beq 
t_{\nu}\sim \left(\sum_{i=1}^4W_i^+\frac1{(\Delta^{\nu}_{i})^2}+W_4^+\frac1{(\Delta^{\nu}_{3})^2}-
2W_4^+\frac1{(\Delta^{\nu}_{3}\Delta^{\nu}_{4})^2}\right)^{-1},
\eeq
where 
the parameters 
\beq
\frac1{\De^{\nu}_i}\equiv\left|\sum_{\alpha=1}^4p_{i\alpha}\sqrt{|\lambda_{\alpha}|}\right|\;\;\;(i=1,..,4), 
\eeq
must fulfill the strong inequalities 
\beq
\frac1{\De^{\nu}_i}\ll 1 \;\;\;\;\;(i=1,..,4),   
\eeq
for the kinetic equation to have the Fokker-Planck form with respect to variables $\nu_1,..,\nu_4$.

Thus, we have 
\beq 
\frac{t_{\xi}}{t_{\nu}}\sim \frac1{2}\frac{\sum_{i=1}^4W_i^+\frac1{(\Delta^{\nu}_{i})^2}+W_4^+
\frac1{(\Delta^{\nu}_{3})^2}-2W_4^+\frac1{\Delta^{\nu}_{3}\Delta^{\nu}_{4}}}
{\sum_{i=1}^3\frac{k_i+1}{k_i}W_i^+\alpha_i^2+ W_4^+\alpha_4^2}\ll 1. 
\eeq
This strong inequality expresses the hierarchy of time scales which has allowed us to  identify the thermal
relaxation stage. During this stage the distribution of  nuclei with respect to the variable $\xi$ approaches
the quasi-equilibrium  distribution, while the distribution with  respect to $\nu_1,\nu_2,\nu_3$, and $\nu_4$
practically does not change.

The quasiequilibrium distribution is an eigenfunction of the principal operator of governing equation (23)
with zero eigenvalue. Therefore, as  follows from eq.(38), the operators of the first four terms on the RHS of
eq.(23) also become important at the end of the stage of thermal relaxation.

\section{Numerical evaluations}


For a numerical illustration of our model, 
we carried out calculations for the concurrent nucleation and chemical aging of AHHO aerosols 
in the air  containing the vapors of three condensable components -- water, $2-$methylglyceric acid
(C$_4$H$_8$O$_4$, as a representative of hydrophilic organics in air), 
and $3-$methyl$-4-$hydroxy-benzoic acid (C$_8$H$_8$O$_3$, as a representative of hydrophobic organics in air),
as well as non-condensable nitrogen oxide, hydroxyl radicals, oxygen, and nitrogen dioxide 
(components 1,2,3,5,6,7, and 8, respectively).   
The non-condensable air components thus played also the role of the carrier (passive) gas in the system. 

Besides the air temperature $T=293.15$ K, the atmospheric conditions were  specified by the  saturation ratios
of vapors of water $\zeta_1$, $2-$methylglyceric acid $\zeta_2=$, and  $3-$methyl-4-hydroxy-benzoic acid
$\zeta_3$, and by  the analogous parameters of noncondensable species assumed to be fixed,  
$\zeta_5=\zeta_6=\zeta_7=\zeta_8=1.001$.  The saturation ratio of water vapor was varied. 

According to Couvidat {\em et al.},$^{55}$ the molecules of $2-$methylglyceric acid can be considered to be 
hydrophilic, whereas  $3-$methyl-4-hydroxy-benzoic acid molecules are hydrophobic.$^{}$   The latter will be
mostly located at the aerosol surface, with the methyl groups -CH$_3$  exposed to the air. Thus, one can
consider the abstraction of the H-atom from the methyl group of a $3-$methyl-4-hydroxy-benzoic acid molecule
as reaction (1),  and identify the radical R\hspace{-0.01cm}{\rad} in eqs.(1)-(3)    and component 4 as the
radicals  
$$\mbox{-CH$_2$-C$_6$H$_3$-OH-COOH},\;\;\mbox{and}\;\;\mbox{-OCH$_2$-C$_6$H$_3$-OH-COOH}, $$  respectively.
Thus, the solution in droplets can be treated as a mixture of functional groups with all   relevant parameters
available in the tables of UNIFAC method for activity coefficients.$^{56-58}$ 

The effect of the droplet surface tension on condensation/nucleation phenomena has been well
investigated.$^{46}$ Aiming mainly at the qualitative sensitivity studies of the thermal relaxation process
with respect to the aggregate equilibrium constant $K_{\mbox{\tiny eq}}$, one can conjecture that the effect
of radicals R$_4$ (resulting from the hydrophobic-to-hydrophilic conversion of $3-$methyl$-4-$hydroxy-benzoic
acid) on the surface tension will be roughly similar to the effect of a hydrophilic component on the surface
tension of its aqueous solutions. Taking this into consideration, we have modeled the surface tension 
$\sigma^{\alpha\beta}(\chi_1,\chi_2,\chi_3)$ of the four-component solution ``water/$2-$methylglyceric
acid/$3-$methyl$-4-$hydroxy-benzoic acid/radical R$_4$ with the surface tension of a model ternary solution 
``water/hydrophilic solute (which would represent $2-$methylglyceric acid and radicals R$_4$ combined)
/hydrophobic solute (which would represent $3-$methyl$-4-$hydroxy-benzoic acid)". As such, we chose the
solution of water, $n-$pentyl acetate (surrogate hydrophobic solute), and methanol (surrogate hydrophilic
solute). An analytical expression for its surface tension $\widetilde{\sigma}$ as a function of its
composition was obtained by Santos {\em et al.}$^{59}$ (see refs.38 and 39 for more details). 

The rate constants of forward reactions in sequence (1)-(3) can be roughly estimated to equal their  gas-phase
analogs, but there are no data on the rate constants of corresponding backward reactions.  Thus, in the
function $F=F(\nu_1,..,\nu_4)$ the aggregate  equilibrium constant $K_{\mbox{\tiny eq}}$ of sequence (1)-(3)
has to be  considered as an adjustable parameter. 


The heat capacities of air $c_g\approx 4.01$ and pure water vapor $c_1\approx 4.54$ were determined   by
linearly extrapolating  data in  {\em CRC Handbook of Chemistry and Physics}$^{60}$ and  with the help  of
formulas given in {\em Thermophysical  Properties of Matter}.$^{61}$ The heat capacity $c_2$ of 
$2-$methylglyceric acid vapor was roughly approximated by that of the gaseous  propylbenzene,$^{62}$ thus 
setting $c_2\approx 17.34$, whereas the heat capacity $c_3$ of $3-$methyl$-4-$hydroxy-benzoic acid vapor was 
assumed to be roughly equal to $c_3=\frac{m_3}{m_2}\times c_2\approx 22.08$.  To estimate the heat capacity 
$c\nu$ of a four component nucleus, we assumed that it can be expected to  be similar to the heat  capacity of
a droplet of an aqueous binary solution of some heavy organic compound with the total number of  molecules
equal to $\nu_c$ and the mole fraction of the organic compound equal to $\chi_{2c}+\chi_{3c}+\chi_{4c}$.  We
used the data for the 
the binary solution of water--glycerol$^{63}$ at 
293.15 K and appropriate glycerol mole fraction $\chi_{2c}+\chi_{3c}+\chi_{4c}$.

The equilibrium vapor pressure and latent heat of condensation/evaporation of pure water were obtained 
by linearly interpolating  data in  {\em CRC Handbook of Chemistry and Physics},$^{60}$ with 
$n_{1\infty}=5.78\times 10^17$ cm$^{-3}$ and  $\beta_1=18.14$. The equilibrium vapor pressures and latent
heats of condensation/evaporation of  pure  $2-$methylglyceric acid and $3-$methyl$-4-$hydroxy-benzoic acid
were evaluated by averaging their values  (two for each quantity, one from ref.48 and one from the web-site
http://www.chemspider.com), so that  $n_{2\infty}=7.0\times 10^12$ cm$^{-3}$, $\beta_2=23.23$ and
$n_{3\infty}=5.4\times 10^12$ cm$^{-3}$,  $\beta_3=22.70$.  

Although the exothermicity of  the gas-phase analogs of reactions (1)-(3) is well-known, we were unable to
find data on the enthalpy of reactions (1)-(3) in the case where the hydrophobic molecule HR is that of
$3-$methyl-4-hydroxy-benzoic acid. Taking into account data provided in ref.31 on the enthalpy of similar
reactions and aiming at only rough,  qualitative numerical estimates, we thus assumed the aggregate enthalpy
$\De H_{}$  of the sequence of reactions (1)-(3) to be about $40$ kcal/mol, or $\beta_4=72.15$.  


Since there exist virtually no theoretical nor experimental data on the thermal accommodation and sticking 
coefficients, the calculations were carried out assuming 
$\alpha_{ci}=1,\;\;\alpha_{ti}=\alpha_g=1\;\,(i=1,2,3)$. 
To illustrate the typical dependence of theoretical predictions on the parameters of the air wherein an
ansemble  of model AHHO aerosols is evolving via nucleation and chemical aging, we evaluated the thermal
relaxation time $t_{\xi}$ and the characteristic time of change of the (four-dimensional) size distribution of
such aerosols  $t_{\nu}$ for  various vapor saturation ratios of water ($\zeta_1$), hydrophilic organic
($\zeta_2$), and hydrophobic organic ($\zeta_3$), considering also several values of the aggregate equilibrium
constant $K_{\mbox{\tiny equ}}$.  Some of the results of calculations are presented in Figures 1 and 2 and
Table 1. 

Figure 1a presents the typical dependence of the thermal relaxation time $t_{\xi}$ on $K_{\mbox{\tiny eq}}$,
with the latter changing in the range from $K_{\mbox{\tiny eq}}=1$ (weak chemical aging) to  
$K_{\mbox{\tiny eq}}=5$  (intensive hydrophobic-to-hydrophilic conversion), 
for five saturation ratios of water vapor in the air: the solid curve is for $\zeta_1=0.15$, the long-dashed
curve for $\zeta_1=0.14$, dash-dotted line for $\zeta_1=0.13$, short-dashed curve for $\zeta_1=0.12$,
and dotted curve for $\zeta_1=0.11$.  
In Figure 2b, the typical dependence of $t_{\xi}$ is plotted as a function of 
$\zeta_1$ in the range from $\zeta_1=0.11$ to $\zeta_2=0.15$ 
at five values of $K_{\mbox{\tiny eq}}$: the solid curve is for $K_{\mbox{\tiny eq}}=0.15$, the long-dashed
curve for $K_{\mbox{\tiny eq}}=0.14$, dash-dotted line for $K_{\mbox{\tiny eq}}=0.13$, 
short-dashed curve for $K_{\mbox{\tiny eq}}=0.12$,
and dotted curve for $K_{\mbox{\tiny eq}}=0.11$. 
All results in Figure 1 are for $T_0=293.15$ K, $\zeta_2=0.01$, and $\zeta_3=0.3$ (the parameters of
noncondensable species are specified above). 

Figure 2 presents the typical dependence of the thermal relaxation time $t_{\xi}$ on the saturation ratio
$\zeta_2$ of the hydrophilic organic vapor   at a fixed saturation ratio $\zeta_3=0$ of the hydrophobic
organic vapor (Figure 2a) and on  the saturation ratio $\zeta_3$ of the hydrophobic organic vapor   at a fixed
saturation ratio $\zeta_2=0.01$ of the hydrophilic organic vapor (Figure 2b).  All results in both Figure 2
are for $T_0=293.15$ K, $K_{\mbox{\tiny eq}}=3$, and $\zeta_1=0.13$ (the parameters of noncondensable
species are specified above). 

As clear from Figures 1 and 2, the thermal relaxation time $t_{\xi}$ monotonically decreases with increasing
saturation ratio $\zeta_i\;\;(i=1,2,3)$ of each condensable component of the air. It is also monotonically
decreases with increasing equilibrium constant $K_{\mbox{\tiny eq}}=3$ of the sequence of reactions (1)-(3).
Thus, one can conclude that the quasi-equilibrium distribution of an ensemble of AHHO droplets with respect
to their temperatures is reached
faster in more metastable vapor mixtures (with higher saturation ratios of vapors of water and hydrophilic and
hydrophobic organics) and when the equilibrium of the 
sequence of chemical aging reactions (1)-(3) is shifted more towards products. These results indicate that the chemical aging of aqueous
organic aerosols significantly enhances the hierarchy of time scales in the evolution of the droplet
distribution function. Since the mechanism of  chemical aging of aqueous organic aerosols (i.e., the sequence
of reactions (1)-(3)) strongly favors the products over reagents, our results indicate that the above
presented procedure for finding the distribution function of the ensemble of droplets at the stage of thermal
relaxation is well substantiated at virtually any combination of saturation ratios of air components, both
condensable and inert.

Table 1 presents the characteristic time of change of the (four-dimensional) size distribution of AHHO
aerosols  $t_{\nu}$ and the ratio $t_{\xi}/t_{\nu}$ for various combinations of $\zeta_1,\zeta_2,\zeta_3$, and
$K_{\mbox{\tiny eq}}$.  As evident from this table, the thermal relaxation time remains much smaller (by at
least two orders of magnitude) than the characteristic time $t_{\nu}$ of the size evolution of the
distribution of droplets.   Thus, one can expect that the hierarchy of time scales in the evolution of droplet
distribution (identified on the basis of relative importance of terms on the RHS of eq.(25)), which allowed us
to obtain an analytical solution  (40) of the kinetic equation (25) at the stage of thermal relaxation, exists
for a variety of hydrophilic and hydrophobic organic vapors (participating in nucleation and chemical aging of
organic aerosol) and a wide range of atmospheric conditions. 



\renewcommand{\arraystretch}{0.75}
\begin{table} 
\renewcommand{\baselinestretch}{1}
{\em }
\renewcommand{\baselinestretch}{1}
{\bf Table:} 
The characteristic time of change of the (four-dimensional) size distribution of AHHO
aerosols  $t_{\nu}$ and the ratio $t_{\xi}/t_{\nu}$ for various combinations of $\zeta_1,\zeta_2,\zeta_3$, and
$K_{\mbox{\tiny eq}}$. 
\vspace{4mm} 
\renewcommand{\baselinestretch}{1}
\begin{tabular}{|c|c|c|c|c|c|c|}\hline
\multicolumn{2}{|c|}
{$K_{\mbox{\tiny eq}}$}& $\zeta_1$ & $\zeta_2$ & $\zeta_3$ & $t_{\nu}\;\;(\mu$s) & $t_{\xi}/t_{\nu}$   
\\ \hline
\multicolumn{2}{|c|}{1}& $0.11$ & $0.01$ & $0.3$ & $119.90$ & $0.0001$  
\vspace{-0.3mm}    
\\
\multicolumn{2}{|c|}{ }& $0.13$ & $0.01$ & $0.3$ & $0.73$ & $0.02$  
\vspace{-0.3mm}    
\\
\multicolumn{2}{|c|}{ }& $0.15$ & $0.01$ & $0.3$ & $9.65$ & $0.001$  
\vspace{-0.3mm}    
\\
\multicolumn{2}{|c|}{ }& $0.15$ & $0.019$ & $0.4$ & $0.58$ & $0.02$  
\vspace{-0.3mm}    
\\
\multicolumn{2}{|c|}{ }& $0.15$ & $0.019$ & $0.5$ & $0.38$ & $0.02$ 
\vspace{-0.3mm}    
\\
\multicolumn{2}{|c|}{ }& $0.15$ & $0.019$ & $0.6$ & $0.24$ & $0.03$ 
\vspace{-0.3mm}    
\\
\multicolumn{2}{|c|}{ }& $0.17$ & $0.019$ & $0.4$ & $0.55$ & $0.02$ 
\vspace{-0.3mm}    
\\
\multicolumn{2}{|c|}{ }& $0.19$ & $0.019$ & $0.4$ & $2.66$ & $0.003$ 
\vspace{-0.3mm}    
\\
\hline
\multicolumn{2}{|c|}{3}& $0.11$ & $0.01$ & $0.3$ & $11.55$ & $0.0007$     
\\ 
\multicolumn{2}{|c|}{}& $0.13$ & $0.01$ & $0.3$ & $8.25$ & $0.0009$     
\\ 
\multicolumn{2}{|c|}{}& $0.13$ & $0.01$ & $0.36$ & $5\times 10^4$ & $10^{-6}$     
\\ 
\multicolumn{2}{|c|}{}& $0.13$ & $0.01$ & $0.42$ & $0.17$ & $0.04$     
\\ 
\multicolumn{2}{|c|}{}& $0.13$ & $0.04$ & $0.3$ & $8.25$ & $0.0009$     
\\ 
\multicolumn{2}{|c|}{}& $0.13$ & $0.07$ & $0.3$ & $8.25$ & $0.0009$     
\\ 
\multicolumn{2}{|c|}{}& $0.15$ & $0.01$ & $0.3$ & $9.22$ & $0.0008$     
\\ 
\multicolumn{2}{|c|}{}& $0.15$ & $0.019$ & $0.6$ & $5.13$ & $0.001$     
\\ 

\hline
\multicolumn{2}{|c|}{5}& $0.11$ & $0.01$ & $0.3$ & $0.08$ & $0.08$     
\\ 
\multicolumn{2}{|c|}{}& $0.13$ & $0.01$ & $0.3$ & $0.22$ & $0.03$     
\\ 
\multicolumn{2}{|c|}{}& $0.15$ & $0.01$ & $0.3$ & $5.48$ & $0.001$     
\\ 
\hline
\end{tabular}
\renewcommand{\baselinestretch}{2}
\end{table}

\section{Concluding remarks} 

In the atmosphere, the formation and evolution of secondary aqueous organic aerosols is most likely to  occur
via concurrent nucleation and chemical aging (heterogeneous  chemical reactions on aerosol particles).  
Most of heterogeneous chemical reactions on the aerosol surface can be expected to be exothermic, accompanied
by the release of some enthalpy. 
Therefore, one can assume that during the chemical aging of a liquid organic aerosol, heterogeneous reactions
on its surface are exothermic. Due to the released enthalpy, the   aerosol temperature may deviate up from the
ambient (air) temperature. This  can substantially affect the process of formation and evolution of
organic aerosols.

So far, however, this effect has never been studied, 
whereas other non-isothermal effects (such as the effects of latent heat of condensation/evaporation,
temperature fluctuations, thermal quasi-isolateness of a nascent droplet)  of both unary and multicomponent
vapor-to-liquid phase transitions have been relatively well investigated   (especially thoroughly in the
theory of unary condensation).  In this work, taking account of the deviation of the droplet temperature from
the air temperature  (due to {\em all} these nonisothermal effects) and using the formalism of classical
nucleation theory, we have 
derived a kinetic equation for the distribution 
of an ensemble of aqueous organic aerosols, evolving via nucleation and concomitant chemical aging.

Our kinetic equation governs the temporal evolution of the five-dimensional distribution function  not only in
the case where the latent heats of condensation and the enthalpy of chemical reactions are relatively small,  
per-molecule quantities being much smaller than the rms equilibrium fluctuation of the droplet energy,  but
also  in the case where  they are of the same order of magnitude as the  rms fluctuation of the droplet
energy.  In the former case the kinetic equation  reduces to the canonical form of the five-dimensional
Fokker-Planck equation, whereas in the latter case it goes beyond the framework of the Fokker-Planck equation
with respect to the variable related to the droplet temperature.

We have established that under conditions of the applicability of the capillarity approximation there exists
the hierarchy of the 
time scales in the evolution of the five-dimensional distribution function of droplets. 
This allows one to  identify the stage of thermal relaxation of droplets at which their   distribution
with respect to their temperatures approaches a  quasi-equilibrium Gaussian distribution, while their
distribution with respect to  the numbers of molecules practically does not change. 

As a numerical illustration, we have considered the homogeneous formation (via nucleation and concomitant
chemical aging) of model aqueous hydrophilic/hydrophobic organic (AHHO) aerosols consisting of water,
$2-$methylglyceric acid (as a hydrophilic compound), and  $3-$methyl-4-hydroxy-benzoic acid  (as a hydrophobic
organic compound), in the air containing  the vapors of  these compounds, as well as  typical atmospheric
gaseous species.  Calculations were carried out for  various values of sticking and thermal adaptation
coefficients. 

Numerical evaluations  have shown that, in the model system considered,  the condition of the hierarchy of
time scales is well fulfilled. This means, as first predicted in CNT by Grinin and Kuni (1989), that the
thermal relaxation of the droplet distribution function occurs much faster than its evolution  with respect to
numbers of droplet molecules.  Our estimates also suggest  that the chemical aging of aqueous organic
aerosols may markedly enhance their formation via nucleation and that such an enhancement becomes more
pronounced with decreasing saturation ratio of water vapor, when the nucleation rate decreases.   


At present, it is not possible to make a  comparison between our theoretical predictions and experimental
data,  because even the most modern experimental methods can not provide data on the time  dependence of
the distribution of aqueous organic droplets with respect to the temperature.  Such a comparison will be
eventually necessary when appropriate experimental  data become available. 
 
In order to describe the evolution of the system after the stage of thermal relaxation and to obtain an
expression for the rate of non-isothermal formation of aqueous organic aerosols via concurrent 
nucleation and chemical
aging, it is necessary  to solve the full kinetic equation  taking account of all the terms contributing to
the temporal evolution of the droplet distribution function. This problem will be the object of our further
research. Of course, as long as there are no experimental nor theoretical data on sticking and thermal
accommodation coefficients 
and the aggregate forward reaction
rate and equilibrium constants of the sequence of chemical reactions (involved in aerosol aging), theoretical
predictions will remain uncertain enough. Nevertheless, we would be able to get   approximate magnitudes of
these coefficients by c theoretical predictions and experimental data for the rates of concurrent nucleation
and chemical aging once the experimental ones  become available. 

However, it is already clear that the enthalpy of heterogeneous chemical reactions can have a significant
impact on the formation and evolution of aqueous organic aerosols via nucleation and concomitant chemical
aging.$^{21,39}$ 
Therefore, the proposed approach to developing a non-isothermal theory of this phenomenon can be expected to
improve current computer models for the distribution of such  aerosol particles with respect to their size and
chemical composition; such a distribution constitutes a necessary component of climate models.$^{2,3}$

\newpage

\section*{References} 
\begin{list}{}{\labelwidth 0cm \itemindent-\leftmargin}

\item $(1)$ A. Kantrowitz, 
{\em J. Chem. Phys.}, 1951, {\bf 19}, 1097-1100.
\item $(2)$ J. Feder, K.C. Russel, J. Lothe and G.M. Pound, 
{\em Adv. in Phys.}, 1966, {\bf 15}, 111-178.
\item $(3)$ F.M. Kuni,
{\it Colloid J. USSR}, 1984, {\bf 46(4)}, 602-609.
\item $(4)$ F.M. Kuni,
{\it Colloid J. USSR}, 1985, {\bf 47(2)}, 238-246.
\item $(5)$ A.P. Grinin and F.M. Kuni, 	
{\em Theor. Math. Phys.}, 1989, {\bf 80}, 968-980.
\item $(6)$ A.P. Grinin and F.M. Kuni, 
{\em Vestnik Leningradskogo universiteta. Seriya Fizika, Khimiya} (in Russian), 1989, {\bf 2}, 91-93.
\item $(7)$ A.P. Grinin and F.M. Kuni,
{\it Colloid J. USSR}, 1990, {\bf 52(1)}, 15-22. 
\item $(8)$ F.M. Kuni and A.P. Grinin, 
{\it Colloid J. USSR}, 1990, {\bf 52(1)}, 40-46. 
\item $(9)$ A.P. Grinin and F.M. Kuni,
{\it Colloid J. USSR}, 1990, {\bf 52(2)}, 301-308. 
\item $(10)$ F.M. Kuni and A.P. Grinin, 
{\it Colloid J. USSR}, 1990, {\bf 52(2)}, 229-236. 
{\em Colloid J. USSR}, 1990, {\bf 52(3)}, 383-389.
\item $(11)$ J. C. Barrett and C. F. Clement, 
{\em J. Aerosol Sci.}, 1991, {\bf 22}. 327-335.
\item $(12)$ I.J. Ford, 
nucleation.  
{\em J. Aerosol Sci.}, 1992,  {\bf 23}, 447-455. 
\item $(13)$ A.P. Grinin, F.M. Kuni, and N.P. Feshchenko, 
{\em Theor. Math. Phys.}, 1992, {\bf 93}, 1173-1183.
\item $(14)$ J.C. Barrett, C.F. Clement and I.J. Ford, 
{\em J. Phys.A: Math.Gen.} 1993, {\bf 26}, 529-548.
\item $(15)$ J. C. Barrett, 
{\em J. Phys.A: Math.Gen.}, 1994, {\bf 27}, 5053-5068.
\item $(16)$ F.M. Kuni, A.P. Grinin, and A.K. Shchekin, 
{\em Physica A}, 1998, {\bf 252}, 67-84.
\item $(17)$ M. Lazaridis and Y. Drossinos, 
{\em J. Phys.A: Math. Gen.}, 1997,  {\bf 30}, 3847-3865.
\item $(18)$ Y. S. Djikaev, F. M. Kuni and A. P. Grinin, 
{\it J. Aerosol Sci.}, 1999, {\bf  30}, 265-277.
\item $(19)$ Y. S. Djikaev, J. Teichmann and M. Grmela, 
{\it Physica A}, 1999, {\bf  267}, 322-342.
\item $(20)$ V.B. Kurasov, 
{\it Physica A}, 2000, {\bf 280}, 219-255.
\item $(21)$ Y. S. Djikaev and E. Ruckenstein, 
{\it J. Phys. Chem. Lett.},  2018, {\bf  9}, 5311-5316.
\item $(22)$ T. Hoffmann, J.R. Odum, F. Bowman, D. Collins, D. Klockow, R.C. Flagan, and J.H. Seinfeld, 
{\it J. Atmos. Chem.}, 1997, {\bf 26}, 189-222.
\item $(23)$ J.R. Odum, T.P.W. Jungkamp, R.J. Griffin, R. C. Flagan and J.H. Seinfeld,  
{\it Science}, 1997, {\bf 276}, 96-99.
\item $(24)$ IPCC, {\it Climate Change 2001: The scientific basis}; Intergovernmental Panel on Climate Change; 
Cambridge University Press, Cambridge, U.K.,2001.
\item $(25)$ H. R. Pruppacher and J. D. Klett,  {\it Microphysics of clouds and precipitation}; Kluwer Academic
Publishers, Norwell, 1997.
\item $(26)$ J. H. Seinfeld and S. N. Pandis,  
{\it Atmospheric chemistry and physics: from air pollution to climate
change}, John Wiley \& Sons, New York, 2006.
\item $(27)$ M. Kanakidou, J. H. Seinfeld, S. N. Pandis, I. Barnes, F. J. Dentener, M. C. Facchini,  
R. Van Dingenen, B. Ervens, A. Nenes, C.J. Nielsen, et al. 
{\it Atmos. Chem. Phys.},  2005, {\bf  5}, 1053-1123.
\item $(28)$ C. A. Pope,  
{\it J.Aerosol Med.},  2000, {\bf  13}, 335-354.
\item $(28)$ S.F. van Euden, A. Yeung, K. Quinlam and J.C. Hogg,  
{\it Proc.Am. Thor.Soc.},  2005, {\bf 2}, 61-67.
\item $(30)$ Shelly L. Miller, William W Nazaroff, Jose L. Jimenez, Atze Boerstra, Giorgio Buonanno,
Stephanie J. Dancer, Jarek Kurnitski, Linsey C. Marr, Lidia Morawska, Catherine Noakes
{\it Indoor Air},  2020, {\bf Accepted}, ??-??.
\item $(31)$ G. B. Ellison, A. F. Tuck and V. Vaida, 
{\it J. Geophys. Res.},  1999, {\bf  104}, 11633-11641.
\item $(32)$ P.K. Quinn, D.B. Collins, V.H. Grassian, K.A. Prather and T.S. Bates, 
{\it Chem. Rev.}, 2015, {\bf 115}, 4383-4399.
\item $(33)$ C. R. Ruehl and K. R. Wilson,  
{\it J.Phys.Chem.A.},  2014, {\bf  118}, 3952-3966.
\item $(34)$ M. D. Petters, A. J. Prenni, S. M. Kreidenweis, P. J. DeMott, A. Matsunaga, Y. B. Lim and 
P. J. Ziemann,  
{\it Geophys. Res. Lett.},  2006, {\bf  33},  L24806.
\item $(35)$  Y. Rudich, N. M. Donahue and T. F. Mentel, 
{\it Annu. Rev. Phys. Chem.}, 2007, {\bf  58}, 321-352.
\item $(36)$  Huang, Y.; Wu, S.; Dubey, M.K.; French, N.H.F., 
{\it Atmos. Chem. Phys.} {\bf 2013}, {\it 12}, 6329-6343.
\item $(37)$ Y. S. Djikaev and E. Ruckenstein, 
{\it J. Phys. Chem. A}, 2014, {\bf  118}, 9879-9889.
\item $(38)$ Y. S. Djikaev and E. Ruckenstein, 
{\it J. Phys. Chem. A}, 2018, {\bf  122}, 4322-4337.
\item $(39)$ Y. S. Djikaev and E. Ruckenstein,  
{\it Adv. Colloid Interface Sci.}, 2019, {\bf  265}, 45-67.
\item $(40)$ Y. S. Djikaev and E. Ruckenstein,  {\it Phys. Chem. Chem. Phys.}, 2019, {\bf 21}, 13090-13098. 
\item $(41)$ Y. S. Djikaev and E. Ruckenstein,  {\it Phys. Rev. E}, 2020 {\bf 101}, 062801. 
\item $(42)$ Y. S. Djikaev and E. Ruckenstein,  {\it Phys. Chem. Chem. Phys.}, 2020, {\bf  22}, 17612-17619.
\item $(43)$ F. M. Kuni, A. A. Melikhov, T. Yu. Novozhilova, and I. A. Terentev, 
{\it Theor. Math. Phys.}, 1990, {\bf 83}(2), 530-542. 
\item $(44)$ F. M. Kuni and A. A. Melikhov, 
{\it Theor. Math. Phys.},  1989, {\bf 81(2)}, 1182-1194.
\item $(45)$ J. Lothe and G. M. J. Pound; in {\it Nucleation}, edited by Zettlemoyer, A. C., 
(Marcel-Dekker, New York, 1969). 
\item $(46)$ D. Kaschiev,  {\it Nucleation : basic theory with applications}  
(Butterworth Heinemann, Oxford, Boston, 2000). 
\item $(47)$ E. Ruckenstein and G. Berim, {\it Kinetic theory of nucleation} (CRC, New York, 2016).
\item $(48)$ H. Reiss, The kinetics of phase transitions in binary systems. {\it J. Chem. Phys.} 18 (1950) 840.
\item $(49)$ D. Stauffer, 
Kinetic theory of two-component (``hetero-molecular") nucleation and condensation. 
{\it J. Aerosol Sci.},  1976, {\bf 7}, 319-333.
\item $(50)$  J.O. Hirschfelder, 
{\it J. Chem. Phys.}, 1974, {\bf 61}, 2690-2694.
\item $(51)$  H. Trinkaus, 
{\it Phys. Rev. B}, 1983, {\bf 27}, 7372-7378.
\item $(52)$ A.E. Kuchma, A.K. Shchekin,  
Multicomponent condensation on the nucleation stage.
{\it J. Chem. Phys.} 150 (2019) 054104.  
\item $(53)$ Y.S. Djikaev, E. Ruckenstein,  M. Swihart,  
On the Fokker-Planck
approximation in the kinetic equation of multicomponent classical nucleation theory
{\it Physica A} 585 (2022) 126375; https://doi.org/10.1016/j.physa.2021.126375.  
\item $(54)$ A.P. Grinin and F.M. Kuni, 
{\em  Vestnik Leningradskogo universiteta. Seriya Fizika, Khimiya} (in Russian), 1982, {\bf 22}, 10-14.
\item $(55)$ F. Couvidat, E. Debry, K. Sartelet and C. Seigneur, 
{\it J. Geophys. Res.}, 2012, {\bf  117}, D10304. 
\item $(56)$ A. Fredenslund, R. Jones and J. Prausnitz,  
{\it AIChE J.}, 1975, {\bf 21}, 1086-1099. 
\item $(57)$ A. Fredenslund, J. Gmehling and P. Rasmussen, {\it Vapor-Liquid
Equilibria Using UNIFAC}, Elsevier, Amsterdam, 1977.
\item $(58)$ H. K. Hansen, P. Rasmussen, A. Fredenslund, M. Schiller and J. Gmehling,  
{\it Ind. Eng. Chem. Res.} 1991, {\bf 30(10)}, 2352-2355.
\item $(59)$ B.M.S. Santos, A.G.M. Ferreira and I.M.A. Fonseca, 
{\it Fluid Phase Equilibria},  2003, {\bf 208}, 1-21. 
\item $(60)$ {\em CRC Handbook of Chemistry and Physics}. 75th Edition; Lide, D. R., Ed.; CRC Press: Boca Raton, 1994-1995.
\item $(61)$ {\em Thermophysical Properties of Matter. The TPRC Data Series }
(1970) (Edited
by Touloukian, Y.S.) Vol.6: Specific Heat. Nonmetallic Liquids and Gases.
IFI/Plenum, New York - Washington. 
\item $(62)$ D.R.Stull and F.D.Mayfield, 
{\it Industrial \& Engineering Chemistry}, 1943, {\bf 35}, 639-645.
\item $(63)$ {\em Perry's Chemical Engineers Handbook}. Perry, R. H.; Green, D. W., Eds; 
McGraw Hill Companies, 1999.

\end{list}

\newpage 
\subsection*{Captions} 
to  Figures 1 and 2 of the manuscript {\sc
``Enthalpy effect on the kinetics of concurrent nucleation and 
chemical aging of aqueous organic aerosols: The stage of thermal relaxation"
}  by {\bf  Y. S. Djikaev} and {\bf B. I. Djikkaity}. 
\subsubsection*{}
\vspace{-1.0cm}   
Figure 1. The dependence of the thermal relaxation time $t_{\xi}$ of an ensemble of atmospheric AHHO 
aerosols, evolving via nucleation and concomitant chemical aging, on various parameters of the surrounding 
air:  a) the relaxation time $t_{\xi}$ as a function of 
the aggregate equilibrium constant $K_{\mbox{\tiny eq}}$ at various saturation ratios of the water vapor, 
$\zeta_1$, as indicated in the figure
panel; b) the relaxation time $t_{\xi}$ as a function of 
the saturation ratio of the water vapor  
$\zeta_1$ at various aggregate equilibrium constants $K_{\mbox{\tiny eq}}$, as indicated in the figure
panel. In both panels (a) and (b) $T_0=293.15$ K, $\zeta_2=0.01$, and $\zeta_3=0.3$ (the parameters of
noncondensable species are specified in the text). 
\vspace{0.3cm}\\ 
Figure 2. The dependence of the thermal relaxation time $t_{\xi}$ of an ensemble of atmospheric AHHO 
aerosols, evolving via nucleation and concomitant chemical aging, on the saturation ratios of organic vapors
in the the surrounding 
air:  a) the relaxation time $t_{\xi}$ as a function of 
the saturation ratio $\zeta_2$ of the hydrophilic organic vapor  
at a fixed saturation ratio $\zeta_3=0$ of the hydrophobic organic vapor; 
b) the relaxation time $t_{\xi}$ as a function of 
the saturation ratio $\zeta_3$ of the hydrophobic organic vapor  
at a fixed saturation ratio $\zeta_2=0.01$ of the hydrophobic organic vapor. 
In both panels (a) and (b) $T_0=293.15$ K, $K_{\mbox{\tiny eq}}=3$, and $\zeta_1=0.13$
(the parameters of noncondensable species are specified in the text). 

\newpage
\begin{figure}[htp]\vspace{-1cm}
	      \begin{center}
$$
\begin{array}{c@{\hspace{0.3cm}}c} 
              \leavevmode
      	      \vspace{3.3cm}
	\leavevmode\hbox{a) \vspace{1cm}} &   
\includegraphics[width=8.3cm]{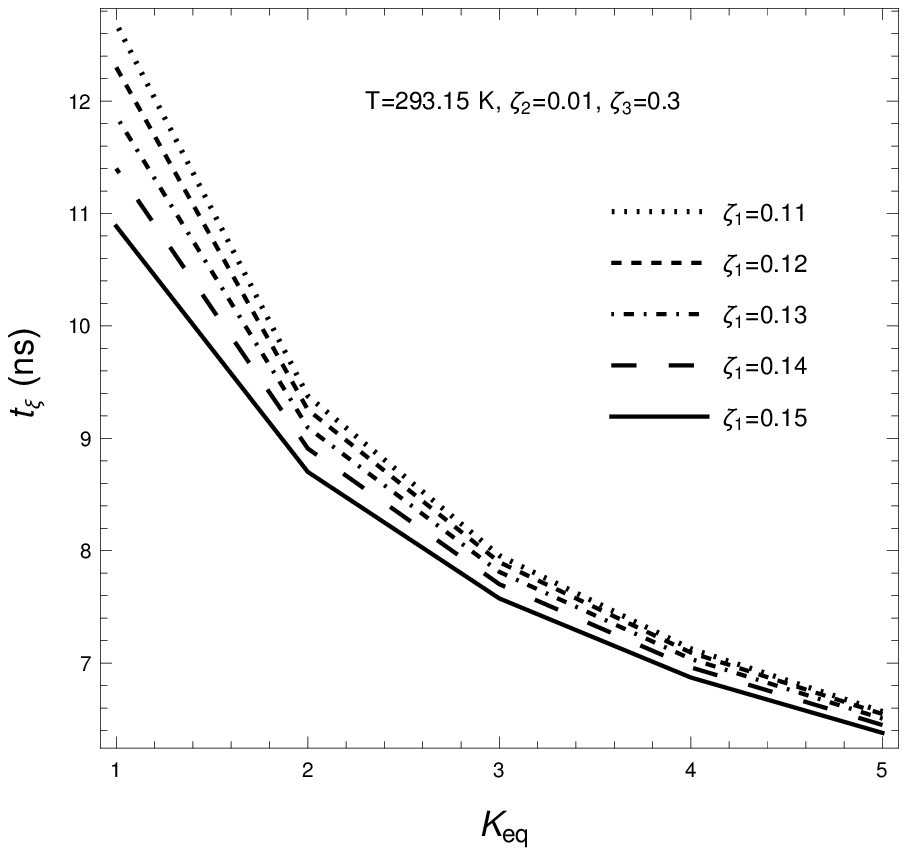}\\ [-0.8cm] 
      	      \vspace{1.0cm}
	\leavevmode\hbox{b) \vspace{1cm}} &  
      	      \vspace{0.0cm}
\includegraphics[width=8.3cm]{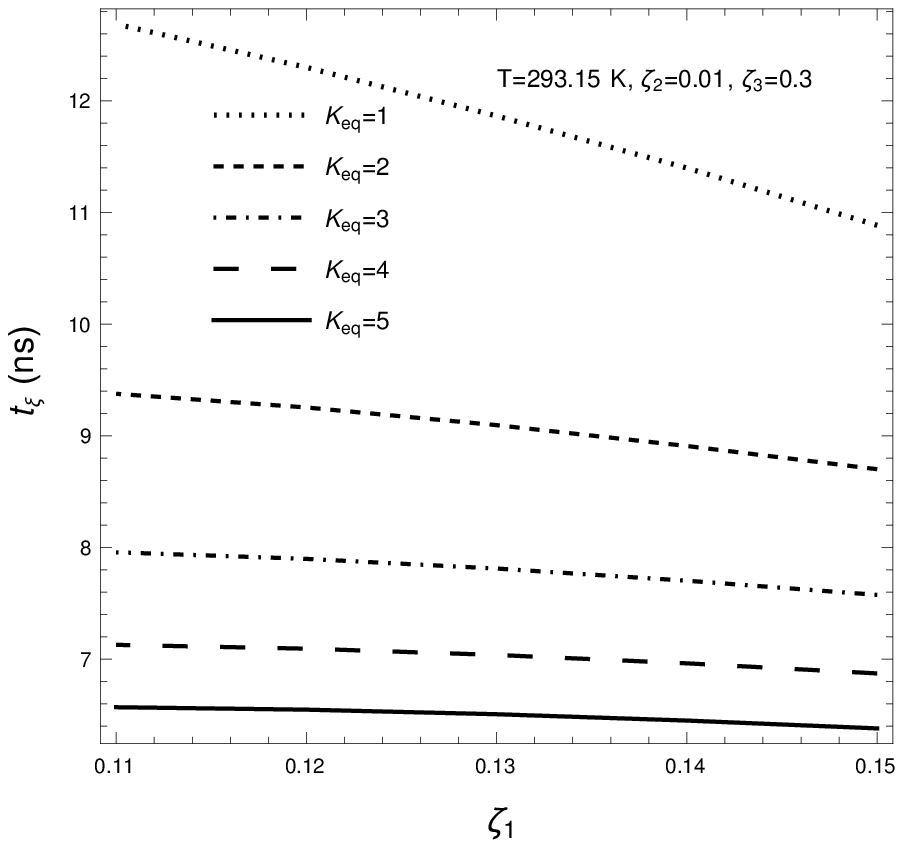}\\ [0.0cm] 
\end{array}  
$$  

	      \end{center} 
            \caption{\small } 
\end{figure}

\newpage
\begin{figure}[htp]\vspace{-1cm}
	      \begin{center}
$$
\begin{array}{c@{\hspace{0.3cm}}c} 
              \leavevmode
      	      \vspace{3.3cm}
	\leavevmode\hbox{a) \vspace{1cm}} &   
\includegraphics[width=8.3cm]{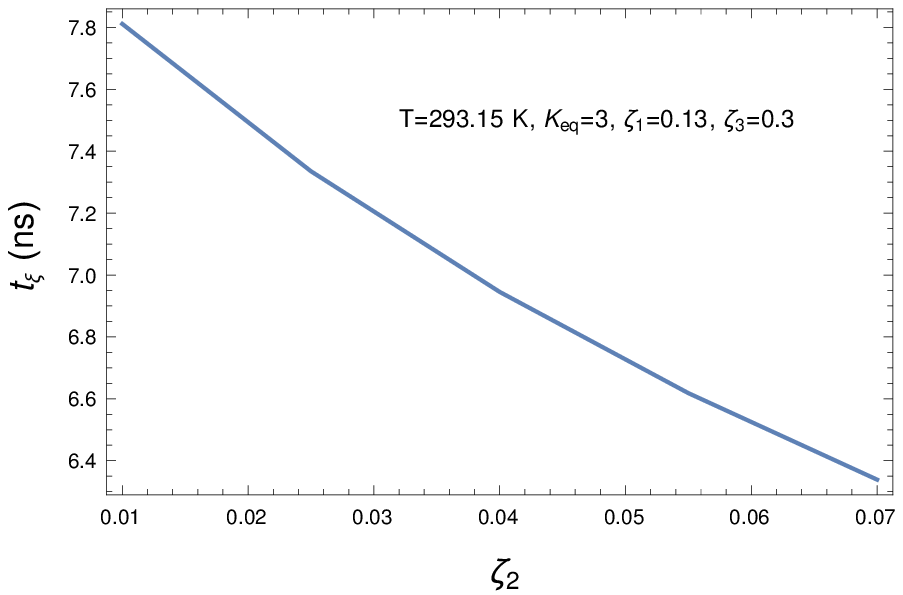}\\ [1.3cm] 
      	      \vspace{1.0cm}
	\leavevmode\hbox{b) \vspace{1cm}} &  
      	      \vspace{0.0cm}
\includegraphics[width=8.3cm]{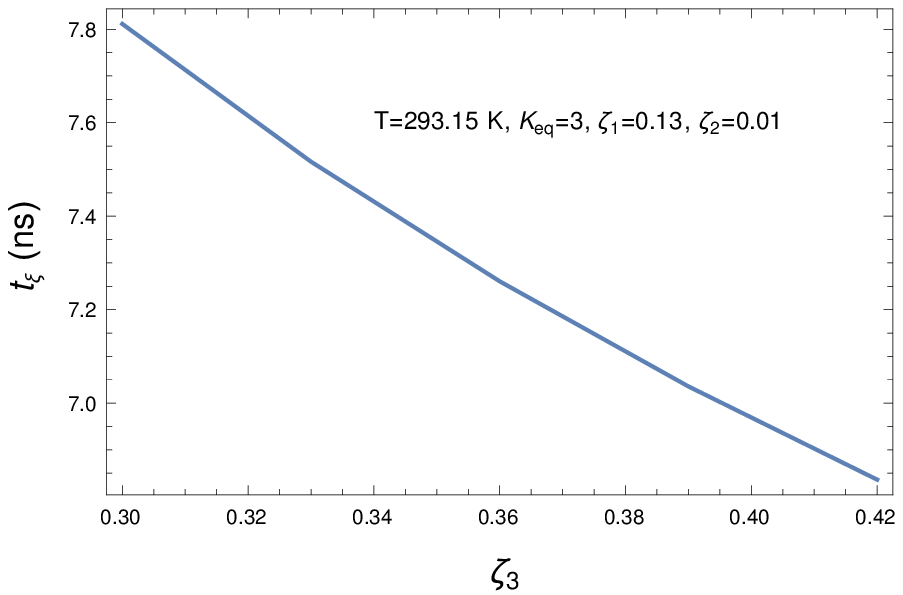}\\ [0.3cm] 
\end{array}  
$$  

	      \end{center} 
            \caption{\small } 
\end{figure}

\end{document}